\documentclass[prb,aps,amsmath,amssymb,preprint]{revtex4}
\usepackage{graphicx}% Include figure files
\usepackage{dcolumn}% Align table columns on decimal point
\usepackage{bm}% bold math
 \begin{document}
 \title{Heterostructure unipolar spin transistors}
\author  {M. E. Flatt\'e}
\affiliation {Optical Science and Technology Center and Department of Physics and Astronomy, University of Iowa, Iowa City,
IA 52242}
\author{G. Vignale}
\affiliation{Department of Physics and Astronomy, University of Missouri, Columbia
MO 65211}
 \begin{abstract}
We extend the analogy between charge-based bipolar semiconductor electronics and spin-based unipolar electronics by considering unipolar spin transistors with different equilibrium spin splittings in the emitter, base, and collector.  The current of base majority spin electrons to the collector limits the performance of ``homojunction'' unipolar spin transistors, in which the emitter, base, and collector all are made from the same magnetic material. This current is very similar in origin to the current of base majority carriers to the emitter in homojunction bipolar junction transistors. The current in bipolar junction transistors can be reduced or nearly eliminated through the use of a wide band gap emitter. We find that the choice of a collector material with a larger equilibrium spin splitting than the base will similarly improve the device performance of a unipolar spin transistor. We also find that a graded variation in the base spin splitting introduces an effective drift field that accelerates minority carriers through the base towards the collector.
 \end{abstract}
 \maketitle
 %\vfill\eject

%%%

Semiconductor spin electronics provides the promise of integrating the non-volatility of metallic magnetoelectronics with the gain properties of semiconductor charge electronics\cite{Spintronicsbook,Ziese}. Semiconductor spin analogs of field effect transistors (spin-FET's\cite{DattaDas,Voskoboynikov:2000,deAndradaeSilva:1999,Koga:PRL2002,Ting:2002,Hall:2003}) and junction transistors\cite{FV,FA,FS1,FS2} have been proposed, although the desired materials properties needed for these devices have yet to be demonstrated. Rapid progress is underway, however, both in the discovery of new ferromagnetic semiconductor materials\cite{GaMnN,TCO,ZnCrTe} and in the improvement of the Curie temperatures of already known ferromagnetic semiconductors\cite{UK,Samarth,Tanaka}. Thus a continued effort is warranted to further develop and improve device designs based on such materials.

\begin{figure}
\includegraphics[width=8cm]{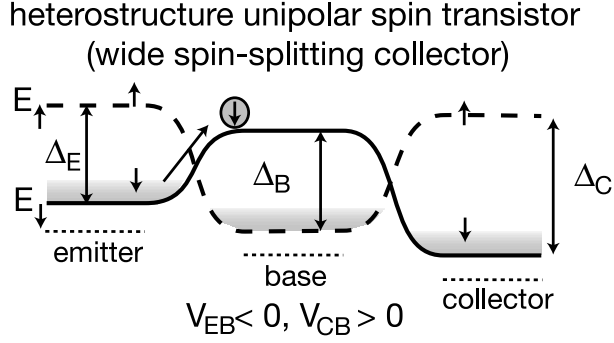}
\caption{
Band edge diagram for a heterostructure unipolar transistor with a wide spin splitting collector. The spin splitting of the emitter is $\Delta_E$, the base is $\Delta_B$, and the collector is $\Delta_C$. The band edges are shown for $\Delta_C>\Delta_B$, which is a good choice to reduce the spin-up current from the base to the collector.  Solid lines are the band edges for spin-down carriers and dashed lines are the band edges for spin-up carriers.  The dotted lines indicate the chemical potential in each region.   
}
\label{diagram}
\end{figure}

%%%
In recent work\cite{FV} we emphasized an analogy between spin-based unipolar junction electronics and bipolar charge electronics. In spin-based unipolar electronics the spin-up and spin-down carriers from a single band play the role of majority and minority carriers ordinarily taken by conduction electrons and valence holes in bipolar devices. The building-block spin device in this approach is the spin diode\cite{FV}, in which two similarly-doped semiconductor regions of opposite magnetization are placed in electrical contact; this situation naturally forms at a 180$^{\rm o}$ domain wall. In this  spin diode, majority (minority) carriers on one side of the device are spin-down (spin-up) electrons and on the other side of the device are spin-up (spin-down) electrons. Under bias the charge current is not rectified, but the spin current is. When two such devices are placed back-to-back in a transistor geometry they amplify charge current in a similar way to bipolar junction transistors. The schematic unipolar spin transistor geometry is shown in Fig.~\ref{diagram}. For $\Delta_E= \Delta_B=\Delta_C$ this diagram is the same as Fig.~2 of Ref.~\onlinecite{FV}.  Such devices can play a role in the design of reprogrammable logic elements, magnetic sensing, and nonvolatile memory (as suggested in Ref.~\onlinecite{FV}), and as differential spin current amplifiers\cite{Fdiff}. Here we show that the use of a $\Delta_C>\Delta_B$ improves the spin polarization of the collector current, the transconductance, and the output conductance  of the unipolar spin transistor. In so doing it is also possible to consider larger base dopings to reduce the base resistance and also the base-width dependence on voltage (the Early effect\cite{Early,Sze}). We further find that the use of a graded spin splitting in the base can accelerate minority carriers through the base towards the collector, which improves both the gain and the switching speed.

The equations governing the emitter, base, and collector currents of these transistors are similar to those governing bipolar transistors. The collector current  is
 \begin{eqnarray}
 I_C &=& -{AqJ_{oB} \over {\rm sinh}(W/L_B)}[({\rm e}^{-qV_{EB}/kT}-1) - ({\rm e}^{-qV_{CB}/kT}
  - 1)\cosh(W/L_B)]\nonumber\\
  && - AqJ_{oC}[{\rm e}^{qV_{CB}/kT}-1]\label{collcurr}
 \end{eqnarray}
 and the emitter current is
 \begin{eqnarray}
 I_E &=& -{AqJ_{oB} \over \sinh(W/L_B)}[({\rm e}^{-qV_{EB}/kT}-1)\cosh(W/L_B) -
 ({\rm e}^{-qV_{CB}/kT}-1)]  \nonumber\\
 &&+ AqJ_{oE} [{\rm e}^{qV_{EB}/kT}-1].\label{emitcurr}
 \end{eqnarray}
 The base width is $W$, the emitter and collector areas are $A$, $k$
 is Boltzmann's constant,  $q$ is the magnitude of the electron charge, $T$ is the temperature and $\hbar$ is Planck's
constant. $J_{oB} = D_Bn_{mB}/L_B$, where $D_B$ is the diffusion
 constant in the base, $n_{mB}$ is the equilibrium minority spin carrier density in the base, and $L_B$ is the minority
 spin diffusion length in the base. $J_{oE}$ and $J_{oC}$ are defined similarly using the appropriate quantities for the emitter and collector respectively. The voltage between emitter and base is $V_{EB}<0$, and the voltage between collector and base is $V_{CB}>0$. The base current is
 $I_B = I_E - I_C$ (this is the convention for common-base amplifiers).  When $W/L_B$ is small, $I_B \ll I_C$, which is the
 desired situation for transistor operation (current gain $I_C/I_B\gg 1$). These equations were reported in Ref.~\onlinecite{FV} for the the base, emitter, and collector all constructed from the same material  with the same doping ($J_{oE}=J_{oB}=J_{oC}$).

Except for the different base, emitter, and collector parameters, the assumptions underlying Eqs.~(\ref{collcurr})-(\ref{emitcurr}) remain the same as in Ref.~\onlinecite{FV}. We assume a negligible number of carriers flip their spin as they move across the junctions from emitter to base, or from base to emitter. This is a similar assumption to assuming the recombination current in bipolar transistors can be neglected in the depletion regions, and is essential for the bulk of the voltage drop to occur across the junction regions.  Detailed calculations of spin transport properties across these magnetic interfaces indicate that the no-spin-flip condition can be met\cite{VignalePRL,DVFJAP}.  We also assume\cite{FV} that the Boltzmann approximation for transport is valid, that the minority carrier densities are small compared to majority carrier densities, and that no generation currents exist in the junction regions.  These are similar assumptions to those underlying common bipolar transistors.

We now take a closer look at the transport of carriers of both spin directions through the device.  Transport processes involved in the movement of spin-down carriers from the emitter to the collector behave nearly identically  to those involved in the motion of electrons from the emitter to the collector in $n-p-n$ bipolar transistors.  For both the bipolar junction transistors and the unipolar spin transistors, however, there are also transport processes involving the other species of carrier which can limit the performance of these transistors. In a $p-n$ junction under forward bias the barriers for minority carrier injection of electrons into the $p$ region and for minority carrier injection of holes into the $n$ region are both reduced, and under reverse bias they are both increased. Thus the problematic junction for bipolar transistors is the forward-biased emitter-base junction, which can permit base majority carriers to be injected at high concentration into the emitter. This also makes it problematic to dope the base layer highly; high base doping would otherwise be desirable, for it can reduce the base resistance and also the Early effect\cite{Early,Sze}. The introduction of a wide band gap emitter\cite{Kroemer,Kroemer2} can be used to keep the barrier high for injection of base majority carriers into the emitter.  In a unipolar spin diode, however, the two types of carriers have the same charge.  Thus a bias which reduces the barrier for spin-down electrons to move in one direction will increase the barrier for spin-up electrons to move the other way\cite{FV}. If the emitter chemical potential in a unipolar spin transistor increases, then the barrier for spin-down electrons to move from the emitter to the base is reduced, and in contrast to the case for bipolar transistors the barrier for the spin-up electrons to move from the base to the emitter is {\it increased}. The problematic junction for unipolar spin transistors, therefore, is not the base-emitter junction. Instead it is the base-collector junction, where an increasing barrier for spin-down electrons to move from the collector to the base implies a decreasing barrier for spin-up electrons to move from the base to the collector. This effect manifests in an unusual  ``collector
multiplication factor" $M$, defined as the ratio between the full collector
current $I_C$ and the majority spin-direction charge current $I_{C
\downarrow}$\cite{Sze}. For a homojunction unipolar spin transistor 
 \begin{equation}
 M = 1+\sinh(W/L_B){\rm e}^{q[V_{CB}+V_{EB}]/kT},
 \end{equation}
and is close to $1$ only if $W/L$ is small and $V_{EB} + V_{CB}<0$. Thus unlike a bipolar transistor, where $|V_{EB}|$ is kept small to reduce base majority current into the emitter, and $|V_{CB}|$ is typically large (but under the avalanche threshold), the unipolar spin transistor operates better ($M\sim 1$) if $|V_{EB}|$ is large and $|V_{CB}|$ is small.

The solution to the current of majority base carriers to the emitter in the bipolar transistor suggests an approach to limit the undesirable spin-up current from the base to the collector in a unipolar spin transistor. This  is to introduce a collector with a larger spin splitting than in the base. In such a heterojunction unipolar spin transistor
  \begin{equation}
 M = 1+({J_{oC}}/{J_{oB}})\sinh(W/L_B){\rm e}^{q[V_{CB}+V_{EB}]/kT}.
 \end{equation}
 The new factor $J_{oC}/J_{oB}$ depends simply on the spin splittings through $n_{mC}/n_{mB}$. Thus  
 \begin{equation}
 M = 1+{\rm e}^{-(\Delta_C -\Delta_B)/kT}({D_{C}L_B}/{D_{B}L_C})\sinh(W/L_B){\rm e}^{q[V_{CB}+V_{EB}]/kT}
 \end{equation}
If $\Delta_C$ exceeds $\Delta_B$ by several $kT$ more than $q(V_{CB}+V_{EB})$, then $M$ can be nearly unity, corresponding to an almost entirely spin-polarized collector current of spin-down carriers.

A very recent analysis of homojunction unipolar spin transistors\cite{BC} has suggested that the output conductance and reverse feedback conductance may be high relative to bipolar junction transistors; this is a consequence of the larger probability for base majority carriers to enter the collector in unipolar spin transistors than in bipolar junction transistors.   This analysis, when applied to heterostructure unipolar spin transistors, yields the following results for the small-signal properties. 
The transconductance
\begin{eqnarray}
g_m = \frac{\partial I_C}{\partial V_{EB}}{\Big|}_{V_{EC}}&=& \frac{Aq^2J_{oB}}{kT{\rm sinh}(W/L_B)}[1-{\rm cosh}(W/L_B){\rm e}^{qV_{EC}/kT}]{\rm e}^{-qV_{EB}/kT} - \frac{Aq^2J_{oC}}{kT}{\rm e}^{qV_{CB}/kT}\nonumber\\&\sim&  \frac{Aq^2J_{oB}}{kT{\rm sinh}(W/L_B)}{\rm e}^{-qV_{EB}/kT} - \frac{Aq^2J_{oC}}{kT}{\rm e}^{qV_{CB}/kT}.
\end{eqnarray}
As $V_{EC}<0$, the quantity in the square brackets can be approximated as unity.  Note that the current of base majority carriers to the collector directly reduces the transconductance of the unipolar spin transistor.  The output conductance
\begin{eqnarray}
g_o = \frac{\partial I_C}{\partial V_{EC}}{\Big|}_{V_{EB}}&=&\frac{Aq^2J_{oC}}{kT}{\rm e}^{qV_{CB}/kT} +\frac{Aq^2J_{oB}}{kT}{\rm coth}(W/L_B){\rm e}^{-qV_{CB}/kT} \nonumber\\&\sim& \frac{Aq^2J_{oC}}{kT}{\rm e}^{qV_{CB}/kT},
\end{eqnarray}
and the reverse feedback conductance
\begin{eqnarray}
g_\mu = \frac{\partial I_B}{\partial V_{EC}}{\Big|}_{V_{EB}}&=&-\frac{Aq^2J_{oC}}{kT}{\rm e}^{qV_{CB}/kT} +\frac{Aq^2J_{oB}[1-{\rm cosh}(W/L_B)]}{kT{\rm sinh}(W/L_B)}{\rm e}^{-qV_{CB}/kT} \nonumber\\&\sim&-\frac{Aq^2J_{oC}}{kT}{\rm e}^{qV_{CB}/kT}.
\end{eqnarray}
The input conductance is $g_\pi + g_\mu$, where
\begin{eqnarray}
g_\pi = \frac{\partial I_B}{\partial V_{EB}}{\Big|}_{V_{EC}}&=& \frac{Aq^2J_{oB}[{\rm cosh}(W/L_B) - 1]}{kT{\rm sinh}(W/L_B)}[1 + {\rm e}^{qV_{EC}/kT}]{\rm e}^{-qV_{EB}/kT} \nonumber\\&&\qquad+\frac{Aq^2J_{oC}}{kT}{\rm e}^{qV_{CB}/kT}
+\frac{Aq^2J_{oE}}{kT}{\rm e}^{qV_{EB}/kT}
\nonumber\\&\sim& \frac{Aq^2J_{oB}[{\rm cosh}(W/L_B) - 1]}{kT{\rm sinh}(W/L_B)}{\rm e}^{-qV_{EB}/kT} +\frac{Aq^2J_{oC}}{kT}{\rm e}^{qV_{CB}/kT}.
\end{eqnarray}
In the heterostructure unipolar spin transistor, the quantities which ideally should be large ($g_m$ and $g_\pi$) have terms proportional to $J_{oB}$, and those which should be small ($g_o$ and $g_\mu$), are only proportional to $J_{oC}$. Hence we can dramatically improve the device performance by taking $J_{oC}/J_{oB}\rightarrow 0$. As we found above
\begin{equation}
J_{oC}/J_{oB}\propto n_{oC}/n_{oB} \sim {\rm e}^{-(\Delta_C-\Delta_B)/kT}.
\end{equation}
Thus the choice of a collector region spin splitting that exceeds the base region spin splitting by many $kT$ will significantly reduce the undesirable conductances associated with the homojunction unipolar spin transistor.

\begin{figure}
\includegraphics[width=8cm]{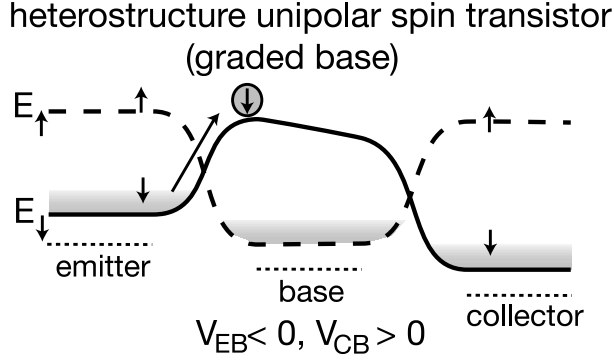}
\caption{
Band edge diagram for the heterostructure unipolar transistor with a graded base. Solid lines are the band edges for spin-down carriers and dashed lines are the band edges for spin-up carriers.  The dotted lines indicate the chemical potential in each region.   The graded $\Delta_B$ in the base produces an effective quasi-electric field that accelerates the spin-down minority carriers in the base towards the collector. It also is easier to combine the graded base of the unipolar spin transistor with the wide spin splitting collector for the narrowest splitting of the base occurs at the interface with the collector.
}
\label{drift}
\end{figure}

We also mention briefly another beneficial design strategy for the heterostructure unipolar spin transistor  motivated by proposals for base band gap grading in heterostructure bipolar transistors\cite{Kroemer,Kroemer2} ---  to grade the spin splitting through the base. As shown in Fig.~\ref{drift} the resulting quasi-electric field will accelerate the spin-down carriers through the base towards the collector. The grading naturally places the smallest  spin splitting on the side of the base nearest the collector. This will enhance the effect of the wide spin splitting collector on reducing the base majority spin current to the collector.  The situation here is different from the bipolar transistor, where the widest-gap region of the base is near the emitter, requiring the use of an even wider-gap material for the emitter region. Analytic expressions for the transistor currents are no longer straightforward with the graded base, but the benefit to transistor performance is clear. Faster minority carrier transport through the base increases gain and decreases switching speed\cite{Sze}.

The device performance advantages of using a heterostructure unipolar spin transistor over a homojunction unipolar spin transistor are another example of the analogy between unipolar spin electronics and bipolar charge electronics emphasized in Ref.~\onlinecite{FV}.  As the two carrier species for unipolar spin transistors have the same charge, the device region which should be modified to improve performance is the collector, not the emitter. Grading of the spin splitting in the base region also will enhance minority carrier transport through the base. We note that these modifications to the spin splitting in the transistor configuration do not affect the alternate (shorted) configuration of the transistor, in which the emitter, base, and collector magnetization are all parallel. Here  the spatial variation of the energy of the minority spin band is irrelevant, as the current will be carried entirely by the  majority carriers.

M.E.F. acknowledges support from DARPA/ARO DAAD19-01-0490. G.V. acknowledges support from NSF Grant Nos. DMR-0313681. We acknowledge discussions with S. Bandyopadhyay and M. Cahay.

\end{document}